\documentclass{article}
\usepackage{amssymb}  % 引入amssymb包
\usepackage{ijcai25}
\usepackage{times}  % DO NOT CHANGE THIS
\usepackage{helvet}  % DO NOT CHANGE THIS
\usepackage{courier}  % DO NOT 
\usepackage{booktabs}  % DO NOT CHANGE THIS
\usepackage[hyphens]{url}  % DO NOT CHANGE THIS
\usepackage{graphicx} % DO NOT CHANGE THIS
\usepackage{natbib}  % DO NOT CHANGE THIS AND DO NOT ADD ANY OPTIONS TO IT
\usepackage{caption} % DO NOT CHANGE THIS AND DO NOT ADD ANY OPTIONS TO IT
\usepackage{fancyhdr}
\usepackage{amsmath}
\usepackage{multirow}
\usepackage{graphicx}
\usepackage{subfigure}
\usepackage{makecell}
\usepackage{algorithm}  
\usepackage{algorithmic}  
\usepackage{enumerate}
\usepackage{xcolor}
\usepackage{times}
\usepackage{soul}
\usepackage{url}
\usepackage[hidelinks]{hyperref}
\usepackage[utf8]{inputenc}
\usepackage{graphicx}
\usepackage{amsmath}
\usepackage{amsthm}
\usepackage{booktabs}
\usepackage{algorithm}
\usepackage{algorithmic}
\usepackage[switch]{lineno}

\title{Towards Automatic Sampling of User Behaviors for \\ Sequential Recommender Systems}

\author{
	Hao Zhang$^1$
	\and
	Mingyue Cheng$^1$\thanks{Corresponding author}\and
	Zhiding Liu$^1$\and
	Junzhe Jiang$^1$\\
	\affiliations
	$^1$State Key Laboratory of Cognitive Intelligence, University of Science and Technology of China\\
	\emails
	\{zh2001, zhiding, jzjiang\}@mail.ustc.edu.cn,
	\{mycheng\}@ustc.edu.cn,
}
\begin{document}
	
	\maketitle
	
	\begin{abstract}{
			Sequential recommender systems (SRS) have gained increasing popularity  due to  their remarkable proficiency in capturing dynamic user preferences. In the current setup of SRS, a common configuration is to  uniformly consider each historical behavior as a positive interaction. However, this setting has the potential to yield sub-optimal performance as each individual item often have a different impact on shaping the user's interests. Hence, in this paper, we propose a novel automatic  sampling framework for sequential recommendation, named AutoSAM, to non-uniformly treat historical behaviors. Specifically, AutoSAM extends the conventional SRS framework by integrating an extra sampler to intelligently discern the skew distribution of the raw input, and then sample informative sub-sets to build more generalizable SRS. To tackle the challenges posed by non-differentiable sampling actions and to introduce multiple decision factors for sampling, we further design a novel reinforcement learning based method to guide the training of the sampler. Furthermore, we theoretically devise multi-objective sampling rewards including \textit{Future Prediction} and \textit{Sequence Perplexity}, and then optimize the whole framework in an end-to-end manner by combining the policy gradient. We conduct extensive experiments on  benchmark recommendation models and four real-world datasets.
			The experimental results demonstrate the effectiveness of the proposed AutoSAM\footnote{\url{https://github.com/zh-ustc/AutoSAM}}.}
		% \footnote{\href{https://anonymous.4open.science/r/AutoSAM-FA40/}{Github: https://anonymous.4open.science/r/AutoSAM-FA40/}}.}
\end{abstract}
\vspace{-0.2in}
\section{Introduction} 
Recommender systems ~\citep{koren2009matrix,liu2018illuminating,zhang2025comprehensive,zhang2024learning} have become crucial tools for information filtering in various online applications, such as e-commerce, advertising, and online videos. Among them, sequential recommender systems (SRS) ~\citep{li2017neural,liu2022one,singer2022sequential,zhang2024learning2} have become increasingly prevalent due to their ability to capture long- and short-term user interests~\cite{cheng2021learning}.

So far, many efforts have been devoted to sequential recommendation, ranging from early matrix factorization and Markov chain based method \citep{rendle2010factorizing} to state-of-the-art deep neural network models~\cite{cheng2022towards}, including recurrent neural networks (RNNs) \citep{hidasi2015session,hidasi2016parallel}, convolutional neural networks (CNNs) \citep{tang2018personalized,yuan2019simple}, and self-attentive models \citep{kang2018self,sun2019bert4rec,li2021lightweight}. Meanwhile, the effectiveness and  efficiency of the generative loss (i.e., auto-regressive loss)~\citep{yuan2019simple} in SRS tasks have been widely demonstrated in these approaches. Despite their remarkable success, these methods often consider from a model perspective while treating all historical behaviors as uniformly positive. In fact, this may lead to sub-optimal performance as items in the sequences are usually of unequal importance on shaping user interests~\citep{zhang2013optimizing}.
% For instance, it is widely recognized that purchased items should contribute more to shaping the user's interests than clicked ones~\citep{wang2021denoising}. Additionally, due to the inherent randomness of user behaviors, even interactions of the same kind can hold significant differences and implications.
% For instance, purchased items should hold more significance than merely clicked ones. 
% Moreover,
% the motivation behind the user behaviors are often very complex and multifaceted. Some interactions on online platforms may be driven by a herd mentality or even the result of mis-clicks, while others may represent the user's true preferences.
For instance, purchased items should hold more significance than merely clicked ones. Moreover, the motivations behind the same action (e.g. clicking) can also be very complex and multifaceted. Interactions on online platforms may sometimes be driven by a herd mentality or could even be the result of mis-clicks, while in other cases, they may accurately represent a user's genuine interests.

% Recently, there have also been some works which attempt to improve the recommender systems from a data perspective~\citep{he2023survey,qin2023learning}. To be specific, SIM \citep{pi2020search} extracts user interests with two cascaded search units to capture the diverse user's long-term interest with target item. Similarly, UBR4CTR \citep{qin2020user} adopts two-stage frameworks to model long-term interests. At the first stage, it generates the query according to the target item for retrieving similar historical behaviors. Then, at the second stage, these items are leveraged to predict the user's interests. To reduce the information loss of the above methods, SDIM \citep{cao2022sampling} proposes a sampling-based end-to-end approach for CTR prediction, which samples from multiple hash functions to gather similar behavior items to the target.
Recently, there have also been some works which attempt to improve the recommender systems from a data perspective~\citep{he2023survey,qin2023learning}. SIM \citep{pi2020search} and UBR4CTR \citep{qin2020user} extracts user interests with search or retrieve units to capture the diverse user's long-term interest with target item. Similarly, SDIM \citep{cao2022sampling} samples from multiple hash functions to gather similar behavior items to the target for CTR prediction.
Though the aforementioned methods have proven effective, they may  suffer from the following limitations.
Firstly, these methods often select items according to the target, which brings challenges to compute scores across all candidate items in parallel and may be only suitable to fine-ranking stage \citep{covington2016deep}. Besides, such a setup may be too strict and fail to consider from a dynamic perspective.
Secondly, these sampling processes are mainly designed for CTR prediction tasks  \citep{guo2017deepfm,wang2017deep}, which are often typically optimized as binary classification problems. As a result, these approaches often cannot be equipped with generative loss in SRS tasks, whose effectiveness and efficiency have been demonstrated in many previous works \citep{kang2018self,yuan2019simple}.

We hold that an ideal solution should be adaptive while maintaining the core training principles of SRS. Hence, to achieve this goal, we propose a general automatic sampling framework, named AutoSAM, to non-uniformly treat historical behaviors for sequential recommendation. To be concrete, an additional sampler layer is first employed to adaptively explore the skew distribution of raw input. Then,
% as shown in Figure \ref{int}, 
informative sub-sets are sampled from the distribution to build more generalizable sequential recommenders. 
% By doing this, it also improves the diversity of training data since various samples are generated in different iteration rounds. 
% \begin{figure}[t]
	%         %\setlength{\belowcaptionskip}{0.5cm}
	%         \centering
	%         \hspace{-10mm}
	%         \includegraphics[width=0.9\columnwidth]{img2/fig1}
	%         \caption{An illustration of a user’s generated samples, items outlined by red boxes stand for the user’s favorite, while the green box means misclicks, etc.}
	%         \vspace{-0.2in}
	%     \label{int} 
	% \end{figure}
In order to overcome the challenges of non-differentiable discrete sampling actions, and 
to introduce multiple decision factors for sampling,
% also make better sampling decisions from a comprehensive perspective, 
we further introduce a novel reinforcement learning based method to guide the training of the sampler due to its flexibility. We believe that a proper decision should not only focus on the target item in the future, but also consider the coherence with the previous context. Along this line, we incorporate the major factors into the reward estimation, including \textit{Future Prediction} and \textit{Sequence Perplexity}. Finally, both the SRS and the sampler can be jointly optimized in an end-to-end manner by combining the policy gradient. As a result, our method exhibits more accurate recommendations than previous approaches, and is generally effective for various sequential recommenders with different backbones. We summarize the contributions as follows:
\begin{itemize} 
	\item[$\bullet$] 
	We propose to sample user behaviors from a non-uniform distribution for SRS task.
	We highlight that the main challenge is to sample historical behaviors dynamically while leveraging the benefits of generative loss.
	\item[$\bullet$] 
	We propose a general automatic sampling framework for sequential recommendation, named AutoSAM, to build generalizable SRS with an additional sample layer. 
	We further introduce a reinforcement learning based method to solve the challenges of non-differentiable actions and design multi-objective rewards to optimize the sampler.
	\item[$\bullet$]
	We conduct extensive experiments on public datasets to show superior recommendation results compared to previous competitive baselines. We also validate the generality of the proposed method and additionally analyze the effectiveness of the sampler.
\end{itemize} 

\begin{figure*}[h]
	\centering
	\vspace{-0.3in}
	\includegraphics[width=0.85\linewidth]{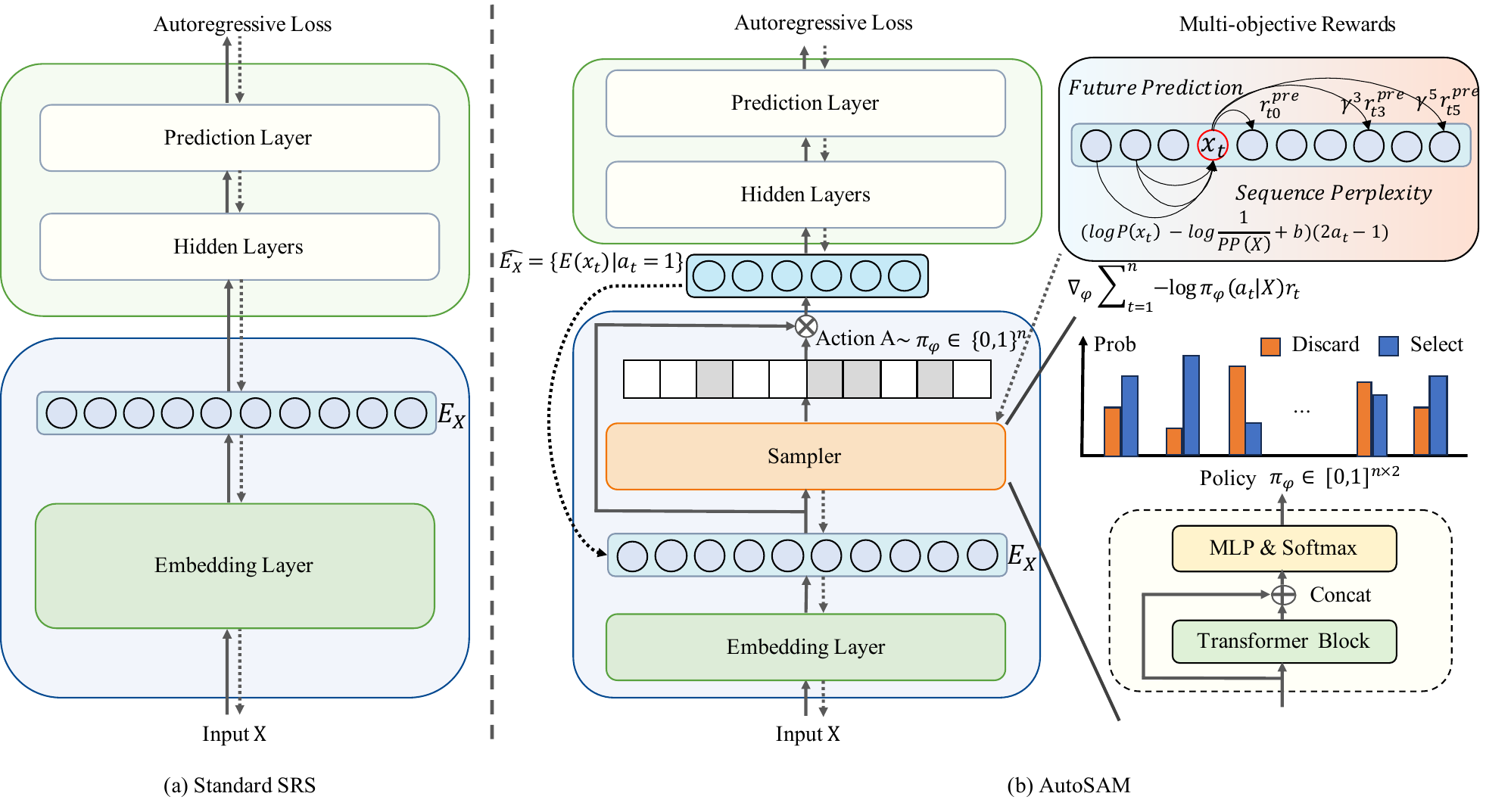}
	% \subfigure[Standard SRS]{
		%     \centering
		%     \includegraphics[width=0.29233\textwidth]{img2/srsa.pdf}
		%     \label{SR}
		% }
	% \hspace{0.1cm}
	% \subfigure[AutoSAM]{ 
		%     \centering
		%     \includegraphics[width=0.67\textwidth]{img2/overviewa.pdf}
		%     \label{AutoSAM}
		% }
	\vspace{-0.15in}
	\caption{(a) is the architecture of the standard sequential recommender system (SRS). (b) is our proposed AutoSAM which augments the traditional sequential recommendation architecture with an additional sampler layer to adaptively explore the skew distribution of raw input. The dashed lines represent gradient backpropagation.}
	\vspace{-0.15in}
	\label{Autosam}
\end{figure*} 
\vspace{-0.15in}
\section{Related Work}
\subsubsection{Sequential Recommendation}
%In order to capture users drifting dynamics of behaviors, sequential recommender systems (SRS) have become ever-increasing popular by learning user and item embeddings from historical behavior sequences.
Sequential recommendation aims to predict users’ future behaviors given their historical interaction data~\cite{luo2024molar,cheng2024empowering}. Early approaches mainly fuse Markov Chain and matrix factorization to capture both long- and short-term item-item transitions \citep{rendle2010factorizing}. Latter, with the success of neural network, recurrent neural network (RNN) methods are widely conducted in sequential recommendation \citep{hidasi2015session,hidasi2016parallel}. Besides, convolutional-based models \citep{tang2018personalized,yuan2019simple}
can also be very effective in modeling sequential behaviors. In addition, we notice that graph neural networks \citep{wu2019session,xu2019graph} have become increasingly prevalent by constructing graph structures from session sequences. And in recent years, self-attention models \citep{kang2018self,sun2019bert4rec,de2021transformers4rec} have shown their promising strengths in the capacity of long-term dependence modeling and parallel computation.
% Furthermore, some methods \citep{xie,qiu2022contrastive} design contrastive learning based module to obtain better representation for sequential behaviors. 
However, these methods focus on the model architectures and treat all behaviors as uniformly positive, may lead to the sub-optimal performance.
% Recently, graph neural networks (GNN) \citep{} have also become increasingly prevalent by constructing graph structures from sessions.
% \subsection{Recommendation Approaches from Data Perspective}
% \vspace{-0.05in}
\subsubsection{User Behavior Sampling} 
As a data-driven technology, recommendation systems
% \citep{rendle2014improving} 
have attracted a series of works that consider from the data perspective in recent years \citep{qin2021world}. To be specific, in SIM \citep{pi2020search}, a two-stage method with a general search unit (GSU) and an exact search unit is proposed to model long-term user behaviors better. Similarly, UBR4CTR \citep{qin2020user} conducts retrieval-based method to achieve the goal. Besides, SDIM  \citep{cao2022sampling} proposed a sampling-based approach which samples similar items to the target by multiple hash functions. And some other works \citep{wang2021denoising,lin2023autodenoise} mainly formulate a denoising task with the aims of filtering irrelevant or noise items. 
Despite effectiveness, these methods are mainly designed for CTR prediction, which may be not suitable to conduct the left-to-right generative loss and sample items dynamically in the SRS task. We also notice some works which attempt to capture more informative patterns between user behaviors sequence by designing advance model architectures. RETR \citep{yao2022recommender} build recommender transformer for sampling the user behavior pathway. And an all-MLP based model is proposed \citep{zhou2022filter} which adopts Fourier transform with learnable filters to alleviate the influence of the noise item. Differently, in this paper, we present a general framework to learn the distributions of the raw inputs adaptively with carefully designed multi-objectives, so as to enhance the SRS with more informative training data. 
\vspace{-0.1in}
\section{Preliminaries}
% Before going into the details, we first introduce the research problem and revisit the sequential recommender systems.
%\vspace{-0.05in}
\subsection{Problem Definition}
Assume that there are item set $\mathcal{I} = \{i_1, i_2, ..., i_{|\mathcal{I}|}\}$ and user set $\mathcal{U} = \{u_1, u_2, ..., u_{|\mathcal{U}|}\}$, we denote each behavior sequence of user $u \in \mathcal{U}$  as $X^u = [x_1^u, x^u_2, ..., x_{n}^u]$, where $x_t^u \in \mathcal{I}$ is the item that $u$ interacted at step $t$ and $n$ is the sequence length. Sequential recommender systems (SRS) aim to predict the item that users might interact with at the next time step. Different from traditional SRS, the main idea behind this work is to sample historical behaviors with a policy $\pi(A^u|X^u)$, where $A^u \in \{select=1, discard=0\}^{n}$, and then utilize  $\hat{X^u}=[x_{k_1}^u,x_{k_2}^u,...,x_{k_{m}}^u]$ to train more generalizable SRS, in which $k$ is the set of time steps about sampled items, $m$ is the length of $\hat{X^u}$.
\subsection{Sequential Recommender System}
As shown in Figure \ref{Autosam}(a), modern SRS contain the components as following. First, each input item $x^u_t$ is mapped into an embedding vector by an embedding lookup operation as:
$ E_t = \zeta_{x_t},$
where $\zeta \in \mathbb{R}^{|\mathcal{I}| \times d}$ is item embedding matrix.
Then, the sequential embedding will be fed into stacked hidden layers to capture the long- and short-term dependence. 
\begin{equation} 
	\begin{aligned}
		\label{layer}
		H^{(l)} &= F^{(l)}(H^{(l-1)}), 1 \leq l \leq L,
	\end{aligned}
\end{equation}
where $L$ stands for the number of layers, $H^{(l)}$ is the output of the $l$-th hidden layer $F^{(l)}$ while $H^{(0)} = E$.
Next, a prediction layer is adopted to generate the distribution $P$  of the user's next preferences for each time step $t$:
\begin{equation} 
	\label{pred}
	P_t = Softmax(H^{(L)}_tW + b),
\end{equation}
in which $W \in \mathbb{R}^{d \times |\mathcal{I}|}$, $b \in \mathbb{R}^{|\mathcal{I}|}$ is the learnable parameters.
Finally, to train the model, the auto-regressive generative loss are employed as optimization objectives, i.e., left-to-right supervision signals. Combining the cross-entropy loss with such an objective, the loss function can be written as:
\begin{equation} 
	\label{rec}
	\mathcal{L}(X;\theta) =  -\sum_{t=1}^{n-1}logP(y_t|X_{\leq t};\theta) = -\sum_{t=1}^{n-1}logP_{t,y_t},
\end{equation}
where $\theta$ is the parameters of the model, $y_t = x_{t+1}$ is the item expect to be predicted at step $t$,
and $n$ is the input length.

\section{AutoSAM: the Proposed Method}
\subsection{Overview of the Framework}
The framework of our proposed AutoSAM is depicted in Figure \ref{Autosam}(b). First, we employ a light-weighted sampler to adaptively learn the non-uniform distribution of the raw input, and then  sample the informative sub-sets from the whole sequence. After that, the sequential recommender systems (SRS) are trained with these higher-quality and diverse samples to gain stronger generalizations. 
Considering the challenges of non-differentiable sampling actions and the necessity to introduce multiple decision factors, we further introduce a novel reinforcement learning based method to optimize the sampler.
Specifically, we treat the sampler as the agent. For each time step $t$ of the user behavior sequence $X$, the sampler takes $X_{\leq t}$ as the current state $s_t$, and outputs an action $a_t \in \{0,1\}$ of the $t$-th item. Then, it observes the carefully designed multi-objective rewards from the environment to update the model via the policy gradient. Finally, the sampler could make optimal decisions by continuously interacting with the environment.
And it is worth mentioning that both the sampler and the SRS can be jointly optimized in an end-to-end manner in our proposed framework.
\subsection{User Behavior Sampler}
We first transform sequence $X$ into embedding $E_X = \zeta_X + P_X$, where $\zeta$ and $P$ are the embedding matrix and position embedding, respectively. We suggest that the sample decisions should be based on both local and global information. Hence, the sampler leverages a Transformer Block with triangular mask matrix to aggregate the global information $X_{\leq t}$ at each step $t$,
and then concatenate it with the local item embedding to generate the sample policy $\pi$ through an MLP layer. The above procedure can be formulated as:
\begin{equation}
	\begin{aligned}
		H &= Transformer(E) \\ 
		S &= ReLU([H||E]W_1+b_1)W_2+b_2, \\
	\end{aligned}
	\label{overview1}
\end{equation}
where $||$ means concatenation, $W_1 \in \mathbb{R}^{2d\times d}, b_1 \in \mathbb{R}^{d},W_2 \in \mathbb{R}^{d\times 2}, b_2 \in \mathbb{R}^{2}$ are learnable parameters.
After that, binary decisions $A \in \{0,1\}^n$ are sampled from the Bernoulli distribution as following:
\begin{equation}
	\begin{aligned}
		& \pi = Softmax(S/\tau) \in \mathbb{R}^{n\times2},\;A \sim \pi \in \{0,1\}^{n},
	\end{aligned}
	\label{overview}
\end{equation}
where $\tau$ is the temperature of the Softmax.
Finally, we can build the sampled sub-set as $\hat{X}=\{x_t|a_t=1\}$.
%\vspace{-0.1in}
\subsection{Multi-Objective Rewards}
Now we discuss how to design our reward, which plays an important role in learning the optimal $\pi$. As mentioned before, traditional SRS often follow the next item prediction task, which can neither update sampler network directly due to the non-differentiable challenge of discrete actions nor control the sample rate. Besides, such single objective may be too strict and one-sided. We argue that the proper decisions should not only focus on the target item but also consider the coherence with preceding context. Along this line, we incorporate the major factors into the reward estimation, including \textit{Future Prediction} and \textit{Sequence Perplexity}.
\subsubsection{Future Prediction}
Generally, items that contribute to predicting the future interactions should be of more opportunities to be selected. Accordingly, we theoretically propose our rewards with consideration of the recommendation loss in Eq. \ref{rec}. Since we train the SRS with the sampled sub-set $\hat{X}$, the objective can be written as:
\begin{equation}
	\small
	\begin{aligned}
		min\;J(\theta,\phi) = min\;\mathbb{E}_X\mathbb{E}_{A\sim \pi_\phi(A|X)}\left[\mathcal{L}(\hat{X};\theta)\right],
		\label{obj}
	\end{aligned}
\end{equation}
where $\theta$ and $\phi$ are the parameters of SRS and sampler, respectively. To optimize this objective, we first transform the prediction loss to an equivalent form:
\begin{equation}
	% \small
	\begin{aligned}
		\mathcal{L}(\hat{X};\theta)
		&= \sum_{t=1}^{m-1}-logP(\hat{y}_t|\hat{X}_{\leq t};\theta)
		\\ &= \sum_{t=1}^{n-1}-a_{t+1}logP(y_t|\{x_{k_j}|1\leq j \leq m-1; k_j\leq t\};\theta)\\ & \overset{\triangle}{=} \sum_{t=1}^{n-1}\mathcal{L}_t(X,A;\theta),
	\end{aligned}
\end{equation}
where $m,n$ is the length of $\hat{X},X$ respectively.
And then derive the gradients with respect to $\phi$, we have:
\begin{equation}
	\label{JJ}
	\begin{aligned}
		&\nabla_\phi J(\theta,\phi) = \mathbb{E}_X\nabla_\phi \sum_{A}\pi_\phi(A|X)\mathcal{L}(\hat{X};\theta) \\
		% &=  \mathbb{E}_X\sum_{A}\pi_\phi(A|X)\nabla_\phi log\pi_\phi(A|X)\mathcal{L}(\hat{X};\theta)\\
		% & = \mathbb{E}_X\sum_{A}\pi_\phi(A|X)\nabla_\phi log\prod_{t=1}^{n}\pi_\phi(a_t|X)\mathcal{L}(\hat{X};\theta)\\
		&= \mathbb{E}_X\mathbb{E}_{A\sim \pi_\phi(A|X)}\nabla_\phi \sum_{t=1}^n [log\pi_\phi(a_t|X)\sum_{t=1}^{n-1}\mathcal{L}_t(X,A;\theta)]\\
		&= \mathbb{E}_X\mathbb{E}_{A\sim \pi_\phi(A|X)}\nabla_\phi \sum_{t=1}^{n} -log\pi_\phi(a_t|X)r^{pre},
	\end{aligned}
\end{equation}
where $r^{pre} = \sum_{t=1}^{n-1}-\mathcal{L}_t(X,A;\theta)$ means the rewards associated to the sampler. 
% More details of mathematical derivations can be found in Appendix. 
In practice, we may make a few changes to it. First, we add a baseline $\mathcal{L}^b$ computed on raw input to $\mathcal{L}$ as it is always positive. Second, we discount the future rewards by a factor $\gamma \in [0,1]$ and ignore the targets whose sample probabilities are lower than $\psi$ to increase stability. Then, the future prediction reward at each time step $t$ can be defined as follows:
\begin{equation}
	r^{pre}_t= \sum_{t'=t}^{n-1}-\gamma^{t'-t}I[\pi_\phi(a_{t'+1}|X)>\psi](\mathcal{L}_{t'}(X,A;\theta)-\mathcal{L}_{t'}^b),
	\label{reward}
\end{equation}
in which $I[ \cdot]$ is the indicator function, $\gamma$ is set to $0.9$ across all the experiments in this work.
\subsubsection{Sequence Perplexity}
% To guide the training of the sampler, the first step is to determining how to evaluate the user sequence. 

In addition to sampling based on future predictions,
it should also be attached great importance to considering the coherence with previous context. 
Inspired by perplexity (PP), which is widely used to evaluate the quality of sentences in Natural Language Processing  (NLP) \citep{liu2016not,dathathri2019plug}  with the definition as:
\begin{equation}
	PP(X) = P(x_1,x_2,...x_n)^{-\frac{1}{n}}= \prod_{t=1}^{n}P(x_t|X_{< t})^{-\frac{1}{n}}.  %= 2^{-\frac{1}{n}\sum_{t=1}^{n}logP(x_t|X_{< t})}
\end{equation}
As words causing high perplexity tend to stray from the context \cite{lin2024not}, similarly, we insist that the items causing high perplexity may lack representativeness. A natural idea is to assign lower sampling probabilities to the items whose predicted probabilities according to historical behaviors are lower than $1/PP(X)$, so as to reduce the perplexity and obtain informative sub-sets in some way. To this end, the sampler should be encouraged to drop $x_t$ ($a_t=0$) while $P(x_t|X_{<t}) < 1/PP(X)$, and select $x_t$ ($a_t=1$) while $P(x_t|X_{<t}) > 1/PP(X)$.

In practice, we employ the SRS to compute the approximate objective at each time step $t$. Besides, a relax factor is further conducted to control the strictness of the sampler.  Above all, the reward can be defined as follow:
\begin{equation}
	r^{pp}_t = \left(logP(x_t|X_{<t}) - log\frac{1}{PP(X)} + b \right) (2a_t-1), 
	\label{rp}
\end{equation}
where $b$ is the relax factor, by which we can indirectly control the global sampling ratio. Generally, the smaller $b$ will lead the sampler to be stricter.
% \vspace{-0.1in}
\subsection{Optimization}
We summarize the rewards with trade-off parameter $\lambda$ and scaling factors $k$ as:
\begin{equation}
	\label{scale}
	r=k(\lambda r^{pre}+(1-\lambda)r^{pp}).
\end{equation}
To further learn the sequential recommendation task, we also derive the gradients according to Eq. \ref{obj} with respect to $\theta$:
\begin{equation}
	\label{theta}
	\nabla_\theta J(\theta,\phi) = \mathbb{E}_X \mathbb{E}_{A\sim \pi_\phi(A|X)}\nabla_\theta \mathcal{L}(\hat{X};\theta).
\end{equation} 
Finally, both the SRS and sampler can be optimized jointly in an end-to-end manner by combining the policy gradient:
\begin{equation}
	\label{update}
	\theta \gets \theta - \alpha_1 
	\nabla_\theta \mathcal{L}(\hat{X};\theta),\;
	\phi \gets \phi + \alpha_2 \nabla_\phi \sum_{t=1}^{n}log\pi_\phi(a_t|X)r_t,
\end{equation}
where $\alpha_1$, $\alpha_2$ are the learning rates of SRS and sampler. Overall,  we present our algorithm in Algorithm \ref{al}.
\subsection{Time and Computation Complexity Analysis}
To evaluate the efficiency of our proposed method for online services,
we analyze the time complexity of AutoSAM during inference in this part. 
% Similar to the training stage, we first sample representative historical behaviors, and then feed them to SRS for modeling the user sequence.
Denote $L$ as the number of layers and $N$ as the sequence length. 
Since we conduct Transformer-based SRS whose time complexity is $\mathcal{O}(LN^2)$, the sampling processing takes $\mathcal{O}(N^2)$ and the sequence modeling can be done within $\mathcal{O}(L(\mathbb{E}[(\sum_{i=1}^N a_i)^2])) = \mathcal{O}( L(\mu^2N^2 + N\sigma^2))$, in which $\mu \in \{0,1\}$ denotes the average sampling rate while $\sigma$ is the variance. Thus, AutoSAM could reduce the time complexity of about $\mathcal{O}((L-\mu^2L-1) N^2)$ by shorten the behavior sequence. Moreover, since  sampling can be regarded as a binary classification problem, we believe the model could be more efficient while maintaining comparable performance by adopting much lighter sampler like RNN or MLP. We will explore such effects in the future work.
\begin{algorithm}[t]
	\caption{Learning the AutoSAM framework} 
	\label{al} 
	\begin{algorithmic}[1]
		%		\REQUIRE $n \geq 0 \vee x \neq 0$ 
		%		\ENSURE $y = x^n$ 
		\STATE Initial SRS parameters $\theta$ and sampler parameters $\phi$.
		\FOR{$Epoch = 1, 2, ... T$}
		\FOR{$u = 1, 2, ... |U|$}
		\STATE $X \gets [x_1^u, x_2^u, ..., x_n^u]$;
		\STATE $\hat{X} \gets [\;]$;
		\STATE Compute Sampling policy $\pi_\phi(A|X) \in [0,1]^{n\times2}$;
		\FOR{$t = 1, 2, ... n$}
		\STATE Sample $a_t$  from $\pi_t$;
		% \IF{$a_t=1$}
		% \STATE Append $x_t$ to $\hat{X}$;
		% \ENDIF
		\STATE \textbf{if} {$a_t=1$} \textbf{then} Append $x_t$ to $\hat{X}$;
		\ENDFOR
		\STATE Compute the recommendation loss  $\mathcal{L}(\hat{X}; \theta) = -\sum_{t=1}^{m-1}logP(\hat{y}_t|\hat{X}_{\leq t};\theta)$ 
		% according to Eq.\ref{rec};
		\STATE Compute reward  $r=k(\lambda r^{pre}+(1-\lambda)r^{pp})$
		% \STATE Compute reward  
		% according to Eq. \ref{reward}, \ref{rp}, \ref{scale};
		% \STATE Update $\theta$ and $\phi$ according to Eq. \ref{update};
		\STATE
		$\theta \gets \theta - \alpha_1 
		\nabla_\theta \mathcal{L}(\hat{X};\theta)$
		\STATE$
		\phi \gets \phi + \alpha_2 \nabla_\phi \sum_{t=1}^{n}log\pi_\phi(a_t|X)r_t$
		\ENDFOR
		\ENDFOR
	\end{algorithmic}  
\end{algorithm}

\vspace{-0.1in}
\section{Experiments}
% 			% In this section, we conduct experiments to answer the following research questions (RQ):
%    % \begin{itemize}
	%    %     \item
	%    % \textbf{RQ-1:} Does AutoSAM outperform the SOTA models? 
	%    % \item \textbf{RQ-2:} Is the proposed framework general useful for various backbone models? 
	% 			% \item \textbf{RQ-3:} Could AutoSAM learn more potential  preferred items of users in the future? 
	% 			% \item \textbf{RQ-4:} What is the impacts of key hyper-parameters to efficiency and effectiveness of the model? 
	%    % \item \textbf{RQ-5:} What are sampled from the historical behaviors? 
	%    %    \end{itemize}
\subsection{Experimental Setup}
\subsubsection{Datasets}
We conduct four real-world datasets from different online platforms as following:
\textbf{Tmall}\footnote{\url{https://tianchi.aliyun.com/dataset/dataDetail?dataId=42}} mainly contains anonymized users' shopping logs of Tmall platform in the past 6 months before and on the "Double 11" day.
\textbf{Alipay}\footnote{\url{https://tianchi.aliyun.com/dataset/dataDetail?dataId=53}} collects huge amount of user data of Alibaba Group and Koubei between July $1^{st}$, 2015 and November $30^{th}$, 2015.
\textbf{Yelp}\footnote{\url{https://www.yelp.com/dataset}} is a public dataset where we obtain each user’s sequential behaviors by ranking its rated items in time order.
\textbf{Amazon Book}\footnote{\url{http://deepyeti.ucsd.edu/jianmo/amazon/index.html}} (abbreviated as \textbf{Amazon}) is selected from Amazon review data. We construct sequential behaviors by using user's rating history. 
Similar to some previous works \citep{zhou2022filter}, we filter inactive users and items with fewer than $c$ interactions, where $c$ is set to 10 for Tmall, Alipay and Amazon while adjusted to 5 for Yelp due to the smaller data scale.  The statics are summarized in Table \ref{data}.
% The statics of these datasets after processing are summarized in Table \ref{data}.

\subsubsection{Compared Methods}
We first compare our method from data perspective with following approaches: 
\textbf{(1) FullSAM} trains normally with full historical behaviors. 
\textbf{(2) FMLP-Rec} \citep{zhou2022filter} conducts Fourier transform with learnable filters to alleviate the influence of noise. \textbf{(3) RETR} \citep{yao2022recommender} builds recommender transformer to sample the user behavior pathway for sequential recommendation. 
\textbf{(4) SDIM} \citep{cao2022sampling} 
%  retrieves relevant historical items
% that are similar to the target item by multiple hash functions to model long-term preference. 
replaces the attention mechanism with multiple hash functions to sample  relevant historical items for modeling long-term preference. 
\textbf{(5) RanSAM}, \textbf{(6) LastSAM},  \textbf{(7) PopSAM}  samples random, last or most popular  behaviors respectively. 
Note that we adopt SASRec \citep{kang2018self} as the sequential recommender system for all sample-based methods due to its effectiveness. 
It is worth mentioning that some other sampling based methods e.g. UBR4CTR, SIM are mainly designed for CTR tasks and typically optimized as a binary classification problem. \textbf{While they sampling behaviors using target item}, \textbf{the sampled user behaviors often depend on what the candidate item is} \citep{cao2022sampling}. As a result, these methods are hard to compute the scores of large candidate set as the million-level setting in this paper.
\begin{table}[t]
	\vspace{-0.1in}
	\centering  \caption{Statics of the used datasets in the experiments.}  
	\vspace{-0.1in}
	\tabcolsep = 0.05cm 
	\resizebox{1\linewidth}{!}{
		\begin{tabular}{cccccc}    \toprule    \ Datasets & \#Num. Users & \#Num. Items & \#Num. Actions & \#Actions/Item &  \#Actions/User  \\    \midrule    \midrule    
			Tmall & 424,170 & 969,426 & 22,010,938 &22.71& 51.89  \\    
			Alipay & 520,064 & 2,076,041 & 20,976,085 & 10.10 & 40.33   \\
			Yelp & 221,397 & 147,376 & 3,523,285 & 23.91 & 15.91 \\    
			Amazon & 724,012 & 1,822,885 & 18,216,875 & 9.99& 25.16 \\    
			\bottomrule    \end{tabular}}
	\label{data}
	\vspace{-0.1in}
\end{table}%

\begin{table}[t]  \centering  
\caption{AutoSAM’s hyper-parameter exploration.}   
\vspace{-0.1in}
	\tabcolsep=0.1cm 
 		\renewcommand{\arraystretch}{1}
	\resizebox{1\columnwidth}{!}{
	 \begin{tabular}{cccccc}    
	 	\toprule
	 	Hyper-parameter & Tuning Range &  Tmall &  Alipay & Yelp & Amazon\\    \midrule    $t$ & $[1.0, 3.0,5.0,7.0,9.0]$ & 5.0 & 5.0 & 5.0 & 3.0 \\   $ b$ & $[-0.5,0.0,0.5,1.0,1.5,2.0]$ & 1.0 & 2.0 & 1.0 & 0.5 
   % \\    $\gamma$ & $[0.8,0.9,1.0]$ & 0.9 & 0.9 & 0.9 & 0.9 
 \\    $k$ & $[2e^{-1},2e^{-2},2e^{-3},2e^{-4},2e^{-5}]$ &$2e^{-3}$& $2e^{-3}$ & $2e^{-2}$ & $2e^{-3}$ \\    $\lambda$ & $[0.25,0.5,0.75]$ &$0.5$& $0.5$ & $0.5$ & $0.5$
   \\    $\psi_0$ & $ [0.2,0.5,0.8] $ & $0.8$ & $0.8$ & $0.8$ & $0.5$ \\    \bottomrule    \end{tabular}
}
        \vspace{-0.2in}
	  \label{t1}\end{table}

\begin{table*}[t]  
	% \vspace{-0.1in}
	\centering  
	\caption{Recommendation performance comparison of different models. The best and the second best performance methods are indicated by bold and underlined fonts. “IMP” denotes the improvements of AutoSAM compared to the best baseline.}   
	\vspace{-0.1in}
	\tabcolsep=0.14cm  
	\renewcommand{\arraystretch}{0.97}
	\resizebox{1\textwidth}{!}{
		% Table generated by Excel2LaTeX from sheet '论文格式'
		\begin{tabular}{ccccc|cccc|cccc|cccc}
			\toprule
			\multirow{2}[2]{*}{Method} & \multicolumn{4}{c}{Tmall} & \multicolumn{4}{c}{Alipay} & \multicolumn{4}{c}{Yelp} & \multicolumn{4}{c}{Amazon} \\
			& N@10 & N@20 & R@10 & R@20 & N@10 & N@20 & R@10 & R@20 & N@10 & N@20 & R@10 & R@20 & N@10 & N@20 & R@10 & R@20 \\
			\midrule
			PopRec & 0.0045 & 0.0053 & 0.0087 & 0.0118 & 0.0136 & 0.0143 & 0.0155 & 0.0182 & 0.0037 & 0.0048 & 0.0066 & 0.0111 & 0.0016 & 0.0018 & 0.0030 & 0.0038 \\
			BPR-MF & 0.0217 & 0.0268 & 0.0402 & 0.0602 & 0.0129 & 0.0142 & 0.0181 & 0.0234 & 0.0173 & 0.0228 & 0.0343 & 0.0564 & 0.0116 & 0.0145 & 0.0212 & 0.0324 \\
			FPMC & 0.0379 & 0.0452 & 0.0683 & 0.0971 & 0.0572 & 0.0632 & 0.0907 & 0.1144 & 0.0210 & 0.0273 & 0.0407 & 0.0657 & \underline{0.0481} & 0.0520 & 0.0662 & 0.0817 \\
			GRU4Rec & 0.0410 & 0.0503 & 0.0765 & 0.1135 & 0.0450 & 0.0518 & 0.0779 & 0.1050 & 0.0213 & 0.0276 & 0.0415 & 0.0666 & 0.0370 & 0.0421 & 0.0581 & 0.0783 \\
			NextitNet & 0.0425 & 0.0510 & 0.0775 & 0.1115 & 0.0498 & 0.0568 & 0.0850 & 0.1124 & 0.0229 & 0.0295 & 0.0447 & 0.0711 & 0.0441 & 0.0493 & 0.0666 & 0.0863 \\
			BERT4Rec & 0.0496 & 0.0595 & 0.0906 & 0.1301 & 0.0485 & 0.0551 & 0.0815 & 0.1078 & 0.0235 & 0.0305 & 0.0460 & 0.0739 & 0.0386 & 0.0441 & 0.0618 & 0.0840 \\
			SR-GNN & 0.0359    & 0.0437 & 0.0664 & 0.0975 & 0.0401 & 0.0459 & 0.0681 & 0.0915 & 0.0199 & 0.0255 & 0.0383 & 0.0608 & 0.0312 & 0.0355 & 0.0485 & 0.0655 \\
			CL4SRec & 0.0516 & 0.0616 & 0.0929 & 0.1322
			& 0.0616 & 0.0694 & 0.1040 & 0.1351 
			& 0.0256 & 0.0322 & 0.0486 & 0.0766 
			& 0.0472 & 0.0528 & 0.0726 & 0.0950 \\
			DuoRec & 0.0526    & 0.0627 & 0.0950 & 0.1350
			& 0.0619 & 0.0698 & 0.1050 & 0.1365
			& 0.0248 &0.0317 & 0.0476  & 0.0752 
			& 0.0478 & 0.0534 & \underline{0.0731} & \underline{0.0954} \\
			\midrule
			FullSAM & 0.0515 & 0.0614 & 0.0927 & 0.1320 & 0.0617 & 0.0695 & 0.1037 & 0.1346 & 0.0245 & 0.0314 & 0.0473 & 0.0746 & 0.0470 & 0.0524 & 0.0718 & 0.0932 \\
			FMLP-Rec &\underline{0.0558} &\underline{0.0665} &\underline{0.1007} &\underline{0.1431} & 0.0630 & 0.0709 & 0.1052 & 0.1362 & \underline{0.0263} & \underline{0.0337} &\underline{0.0506} & \underline{0.0798} & 0.0480 & \underline{0.0535} & 0.0729 & 0.0949 \\
			RETR & 0.0526 & 0.0624 & 0.0942 & 0.1332 & 0.0608 & 0.0688 & 0.1027 & 0.1346 & 0.0250 & 0.0318 & 0.0478 & 0.0751 & 0.0471 & 0.0525 & 0.0711 & 0.0926  \\
			SDIM & 0.0527 & 0.0626 & 0.0939 & 0.1333 & 0.0585 & 0.0685 & 0.0991 & 0.1332 & 0.0241 & 0.0308 & 0.0464 & 0.0734 & 0.0464 & 0.0513 & 0.0701 & 0.0898 \\
			RanSAM & 0.0532 & 0.0633 & 0.0961 & 0.1354 & \underline{0.0633} & \underline{0.0713} & \underline{0.1061} & \underline{0.1380} & 0.0246 & 0.0315 & 0.0475 & 0.0749 & 
			{0.0443}& {0.0500} & {0.0685} & {0.0914} \\
			% \underline{0.0482}& \underline{0.0539} & \underline{0.0740} & \underline{0.0967} \\
			% PopSAM & 0.0516 & 0.0617 & 0.0933 & 0.1337 & 0.0617 & 0.0695 & 0.1037 & 0.1346 & 0.0245 & 0.0314 & 0.0473 & 0.0746 & 0.0470 & 0.0524 & 0.0718 & 0.0932 \\
			LastSAM & 0.0502 & 0.0600 & 0.0904 & 0.1291 & 0.0599 & 0.0667 & 0.0993 & 0.1297 & 0.0234 & 0.0297 & 0.0445 & 0.0698 & 0.0446 & 0.0514 & 0.0696 & 0.0893 \\
			PopSAM &  0.0522 & 0.0623 & 0.0945 & 0.1347  & 0.0590 & 0.0665 & 0.0995 & 0.1292 & 0.0246 & 0.0316 & 0.0474 & 0.0752 & 0.0442 & 0.0496 & 0.0681 & 0.0893 \\
			AutoSAM & \textbf{0.0604} & \textbf{0.0717} & \textbf{0.1084} & \textbf{0.1528} & \textbf{0.0672} & \textbf{0.0753} & \textbf{0.1117} & \textbf{0.1437} & \textbf{0.0272} & \textbf{0.0347} & \textbf{0.0521} & \textbf{0.0817} & \textbf{0.0549} & \textbf{0.0610} & \textbf{0.0831} & \textbf{0.1074} \\
			\midrule
			IMP & 8.24\% & 7.82\% & 7.65\% & 6.78\% & 6.16\% & 5.61\% & 5.28\% & 4.13\% & 3.42\% & 2.97\% & 2.96\% & 2.38\% & 
			14.14\% & 14.02\% & 13.68\% & 12.58\% \\
			% 11.39\% & 13.17\% & 12.30\% & 11.07\% \\
			\bottomrule
			\vspace{-0.1in}
		\end{tabular}%
	} \label{overall}\end{table*}

\begin{figure*}[t]
	\vspace{-0.15in}
	\subfigure[Tmall]{
		\centering
		\includegraphics[width=0.48\linewidth]{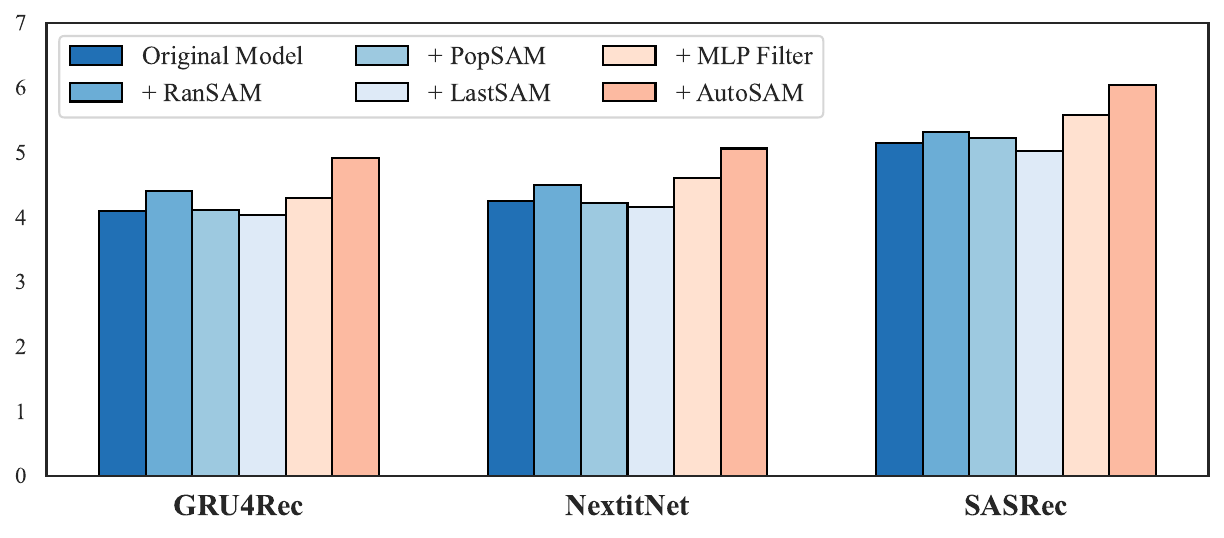}
	}
	\subfigure[Yelp]{
		\centering
		\includegraphics[width=0.48\linewidth]{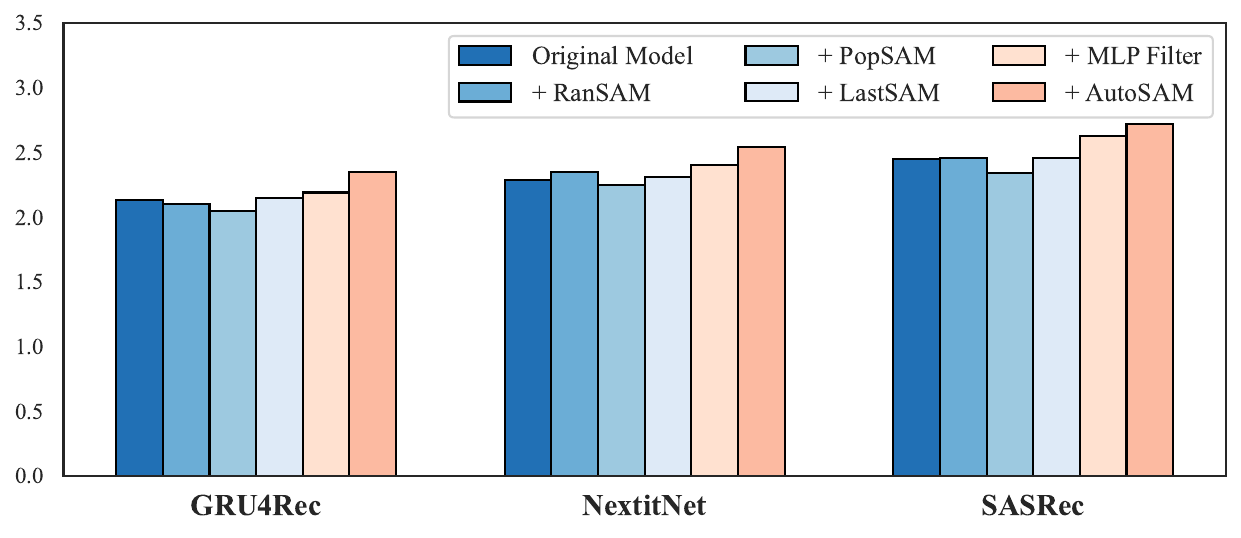}
	}
	\vspace{-0.15in}
	\caption{Performance comparison (NDCG@10) using different backbones on Tmall and Yelp dataset.}
	\label{general1}
	\vspace{-0.2in}
\end{figure*}
Besides, We also compare the models with other architectures:
\textbf{(8) PopRec} is a popularity-based method, in which each user is
recommended according to the popularity.   \textbf{(9) BPR-MF} \citep{rendle2012bpr} is a well-known matrix factorization-based method optimized by Bayesian personalized ranking loss.   \textbf{(10) FPMC} \citep{rendle2010factorizing} combines matrix factorization with the Markov chain to predict the next interaction.   \textbf{(11) GRU4Rec} \citep{hidasi2015session}  is a pioneering attempts by employing recurrent neural network for next item recommendation.   \textbf{(12) NextitNet} \citep{yuan2019simple}  is a CNN-based method to capture long-term and short-term interests. \textbf{(13) BERT4Rec} \citep{sun2019bert4rec} utilizes a bidirectional self-attention network to model user sequence.  
\textbf{(14) SR-GNN} \citep{wu2019session} applies GNN with attention network to model each session.
\textbf{(15) CL4SRec} \citep{xie2022contrastive} utilizes contrastive learning to extract the discrimination information in sequential recommendation. 
\textbf{(16) DuoRec} \citep{qiu2022contrastive} is proposed with a contrastive objective serving as the regularization over sequence representations.

\subsubsection{Hyper-parameter Settings}
We set the batch size of 128 with 10, 000 random negative items per batch, and the embedding size is set to 128 for all methods. In all sampling based methods which employ two-layer SASRec as the backbone, we conduct 4 multi-head self-attention and 2 FFN layers, while the hidden size is set to 256.
% For a fair comparison, the sample rates of RanSAM and LastSAM are set to the same as our well-trained sampler. 
The sample rates of RanSAM, LastSAM and PopSAM are searched from 
$\{0.5,0.6,0.7,0.8,0.9\}$.
We consistently employ Autom as the default optimizer for all recommenders, combined with a learning rate $\alpha_1$ of $1 \times e ^{-3} $. As for our sampler, we conduct SGD with a learning rate $\alpha_2$ of $1 \times e ^{-1} $. 
We use grid search to find the best group of AutoSAM's hyper-parameters as shown in Table \ref{t1}

\subsubsection{Evaluation Metrics}
We evaluate the recommendation performance over all candidate items with \textit{Recall} and  \textit{Normalized Discounted
	Cumulative Gain (NDCG)}. 
%and \textit{Mean Reciprocal Ranking (MRR)}.
The first one is an evaluation of unranked retrieval sets while the other reflects the order of ranked lists. We consider top-k of overall item set for recommendations where $k \in \{10,20\}$ in our experiments. We split the dataset into training, validation and testing sets following the leave-one-out strategy
\citep{cheng2022towards,zhao2022revisiting}.

%      \begin{figure*}[t]
	%      %\vspace{-0.15in}
	% 	\subfigcapskip = -8pt
	% 	\subfigure[N@10, Tmall]{
		% 		\centering
		%   	\includegraphics[width=0.24\linewidth]{img2/tmall_Nrb3.pdf}}
	% 	\hspace{-3mm}
	% 	\subfigcapskip = -8pt
	% 	\subfigure[R@10, Tmall]{
		% 		\centering
		% 		\includegraphics[width=0.24\linewidth]{img2/tmall_Rrb3.pdf}
		% 	}
	% 	\hspace{-3mm}
	% 	\subfigcapskip = -8pt
	% 	\subfigure[N@10, Yelp]{
		% 		\centering
		% 		\includegraphics[width=0.24\linewidth]{img2/yelp_Nrb3.pdf}
		% 	}
	% 	\hspace{-3mm}
	% 	\subfigcapskip = -8pt
	% 	\subfigure[R@10, Yelp]{
		% 		\centering
		% 		\includegraphics[width=0.24\linewidth]{img2/yelp_Rrb3.pdf}
		% 	}
	%       %\vspace{-0.15in}
	% 	\caption{Performance comparison of different
		% 		backbones enhanced by AutoSAM.}
	%       %\vspace{-0.1in}
	% 	\label{general}
	% 	%\Description{des}
	% \end{figure*} 
\subsection{Experimental Performance Analysis}
\vspace{-0.04in}
\subsubsection{Overall Performance.} The result of different models on datasets are shown in Table \ref{overall}.
First of all, we can find that considering from the data perspective instead of treating all items as uniformly positive can benefit the performance. However, SDIM  achieves limited improvement compared to FullSAM. A possible reason is that 
SDIM samples according to the last item maybe too strict to capture the user's rich interests. 
% these methods lost too much information in the process of retrieving or building pathway.
Besides, FMLP-Rec achieves much better performance than traditional SRS probably due to its stronger ability to filter noise information. And we surprisingly find that RanSAM outperforms FullSAM, this suggests that sampling can be regarded as a method of data augmentation in some way, which lead to significant enhancement in generalization capability by increasing the diversity of the training sequence. 
Unfortunately, RanSAM treats each interaction uniformly and may also drop lots of high-quality behaviors, making the improvement limited. 
Different from these baselines, the proposed AutoSAM adaptively explore the skew distribution of the raw input, and then enhance the sequntial recommender with informative samples. Consequently, our method performs best on all datasets, which gains an average improvement over the best baseline of \textbf{7.40\%}, \textbf{7.99\%} on Recall@10, NDCG@10, and  \textbf{6.47\%},  \textbf{7.60\%} on Recall@20, NDCG@20 respectively. 
%		our AutoSAM combines the advantages of above approaches and is more effective.
%		First of all, it is clear that sequential recommendation approaches perform better than non-sequential models which exhibit the importance of sequence dependence.
%		SDIM and RETR achieve only comparable performance as SASRec.
%		A possible reason is that these methods lost too much information in the process of retrieving or building pathway.
%		Besides, FMLP-Rec achieves much better performance than traditional sequential recommenders probably due to its stronger ability in filtering noise information. And we find that RS-SASRec outperforms SASRec, this suggests that sampling can lead to significant improvements in generalization capability. 
%		
%		Finally, our proposed AutoSAM performs consistently better than them on all four datasets. AutoSAM gains an average improvement over the best baseline of \textbf{7.02\%}, \textbf{7.78\%} on Recall@10, NDCG@10, and  \textbf{6.15\%},  \textbf{7.21\%} on Recall@20, NDCG@20 reroundctively.   Different with these baselines, we propose an automatic sampling framework to train the sequential recommenders with informative samples while also keep diversity and variations. As a result,
%		our AutoSAM combines the advantages of above approaches and is more effective.
\vspace{-0.07in}
\subsubsection{Generality Analysis}
% The generality of the method is usually of great value while various models are conducted for different scenarios.
To investigate the general applicability of AutoSAM to other models, we conduct a performance comparison between models using different sampling or filtering methods on various architectures, including RNN-based GRU4Rec \citep{hidasi2015session}, CNN-based NextitNet \citep{yuan2019simple}  and Transformer-based SASRec \citep{kang2018self}. 
It should be noted that since RETR and SDIM are improvements based on attention mechanisms, they are not universally applicable to other architectures. 
Figure \ref{general1} presents the comparison results in terms of test NDCG@10 on Yelp and Tmall datasets. Note that "+ MLP Filter" corresponds to FMLP-Rec. It is evident that all the sequential recommenders demonstrate the most significant improvements through the integration of automatic sampling, highlighting the generality of our AutoSAM for different backbones.
\subsubsection{Ablation Study of Sampling Reward.}
% 		The former part of Eq \ref{reward} implies the benefit of items bringing to the prediction, while the latter means the reward about representative and importance of the target items.  Donate them as $r_{item}$ and $r_{target}$, 
In the left column of Table \ref{ablation},  we analyze the efficacy of each component of the reward. From the results, we can find that removing either of the components decreases the performance. Besides, $r^{pp}$ plays a more important role than $r^{pre}$. A possible reason is that context coherence is more targeted at a single item which could more directly reflect its  representative.
We 	further show the impact of scaling factor $k$ and trade-off parameter $\lambda$ by performing a grid search from \{$2e^{-1}$, $2e^{-2}$, ..., $2e^{-5}$\} and  $\{0,0.25,0.5,0.75,1\}$, respectively. 
The model achieves the best performance at $(k,\lambda) = (2e^{-3},0.5)$.
The results  demonstrate the effectiveness of the incorporation of these two aspects, and it also exhibits the stability of AutoSAM while the different settings of $k$ within the appropriate range bring few influence.
\begin{table}[t]
	\centering
	\caption{The Recall@20 improvement (\%) compared to the original model of users grouped by varying lengths.}
	\tabcolsep=0.5cm   
	\vspace{-0.1in}
	\resizebox{0.9\linewidth}{!}
	{
		\begin{tabular}{ccccc}
			\toprule
			\multirow{2}[2]{*}{Dataset} & \multicolumn{4}{c}{Sequence Length} \\
			& 0$\sim$25 & 26$\sim$50 & 51$\sim$75 & 76$\sim$100 \\
			\midrule
			Yelp & 5.97 & 8.33 & 13.19 & 14.63 \\
			Tmall & 13.45 & 14.58 & 16.56 & 17.40 \\
			\bottomrule
	\end{tabular}}%
	\label{seqlen}%
\end{table}%

\begin{table}[t]  
	\vspace{-0.1in}
	%\begin{minipage}[p]{0.5\textwidth}
	\centering  
	%	\caption{Ablation study of each reward component. Experiments are conducted on Alipay Dataset. Note that we keep the other $\lambda$  at 1e-3 while tuning one of parameters.}    
	\caption{Ablation study of each reward component, scaling factors $k$ and trade-off parameter $\lambda$ on Alipay dataset. "w/o r" means using the random initialized sampler throughout. We fix $\lambda$ as $0.5$ when tuning $k$, and fix $k$ as $2e^{-3}$ when tuning $\lambda$.}   
    \vspace{-0.1in}
	\tabcolsep=0.1cm   
	\resizebox{1\linewidth}{!}
	{
		\begin{tabular}{lcc|ccc|ccc}    \toprule    Method & N@10 & R@10 & $k$ & N@10 & R@10 & $\lambda$ & N@10 & R@10 \\    \midrule    \textbf{AutoSAM}  &  \textbf{0.0672} &  \textbf{0.1117} & $2e^{-1}$ & 0.0663 & 0.1111 & 0.00 & 0.0663 & 0.1109 \\ 
			w/o $r^{pre}$ & 0.0663 & 0.1109 & $2e^{-2}$ & 0.0666 & 0.1113 & 0.25 & 0.0668 & 0.1115 \\    
			w/o $r^{pp}$ & 0.0603 & 0.1016 & \textbf{$2e^{-3}$} & \textbf{0.0672} &  \textbf{0.1117} & \textbf{0.50} & \textbf{0.0672} &  \textbf{0.1117} \\    w/o $r$ & 0.0581 & 0.0984& $2e^{-4}$ & 0.0664 & 0.1109 & 0.75 & 0.0634 & 0.1057 \\ 
			&  & & $2e^{-5}$ &0.0596  &0.0995  & 1.00 & 0.0603 & 0.1016 \\    \bottomrule    \end{tabular}
	} 
	\label{ablation}%\end{table}%
	\vspace{-0.15in}
\end{table}%
\begin{figure}[t]
	\vspace{-0.1in}
	% \hspace{-3mm}
	\subfigure[Tmall]{
		\centering
		\includegraphics[width=0.45\linewidth]{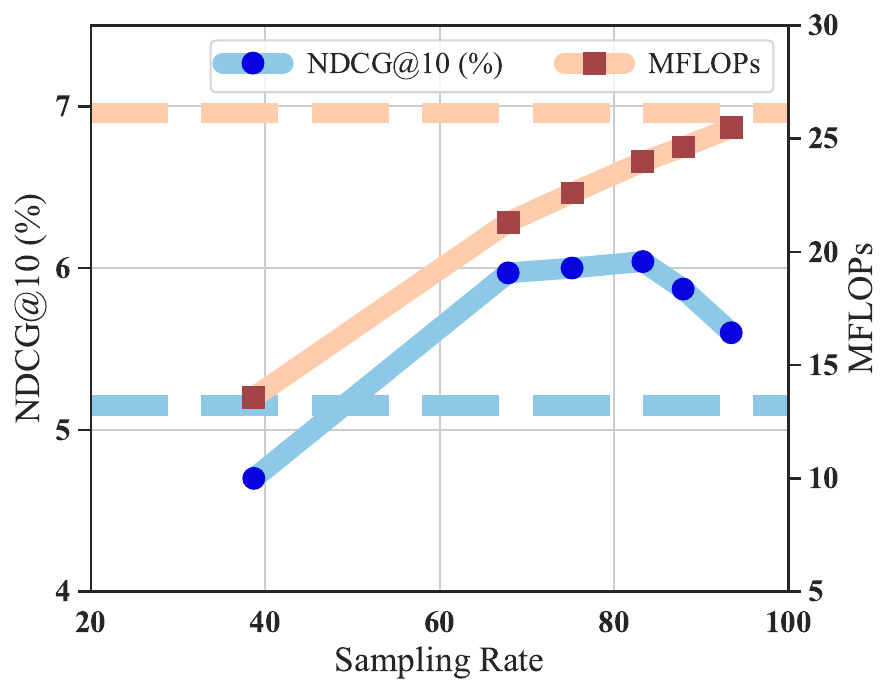}
	}
	% \hspace{-3mm}
	%   \subfigure[Alipay]{
		% 			\centering
		% \includegraphics[width=0.45\linewidth]{img2/autosam_b_alipay.pdf}
		% 		}
	% \hspace{-5mm}
	\subfigure[Yelp]{
		\centering
		\includegraphics[width=0.47\linewidth]{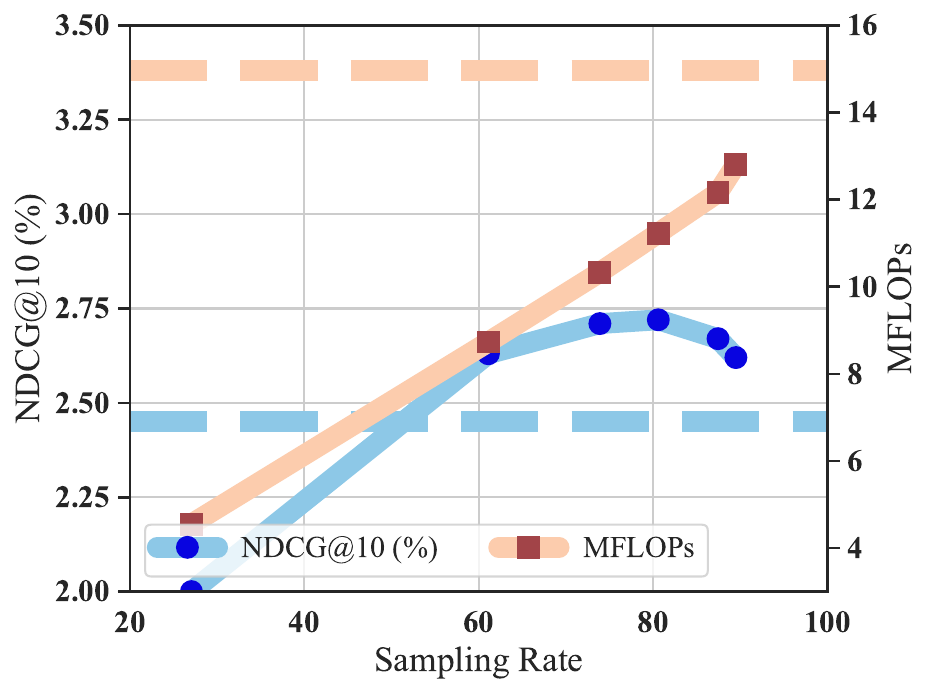}
	}
	% \hspace{-3.5mm}
	%   \subfigure[Amazon]{
		% 			\centering
		% \includegraphics[width=0.45\linewidth]{img2/autosam_b_ama.pdf}
		% 		}
	\vspace{-0.15in}
	\caption{Performance and computation cost w.r.t. sample rates by setting relax factor from $\{-0.5, 0, ..., 2\}$. The dashed line represents the baseline model, i.e., FullSAM.}
	\vspace{-0.15in}
	\label{flop}
	
\end{figure}
% \vspace{-0.1in}
%\subsubsection{Effectiveness of Each Reward Component}

%    \begin{figure}[t]   
	% 				% \subfigcapskip = -5pt
	% 				% \subfigure[Tmall]{
		%     %                 \hspace{-5.2mm}
		% 				% 	\centering
		% 				% 	\includegraphics[width=0.51\linewidth]{img/tmallf1.pdf}
		% 				% }
	% 				% % \subfigcapskip = -5pt
	% 				% \subfigure[Alipay]{
		%     %                 \hspace{-5.5mm}
		% 				% 	\centering
		% 				% 	\includegraphics[width=0.51\linewidth]{img/alif1.pdf}
		% 				% }
	%     %                 \\
	%     %                 %\vspace{-3mm}
	% 				% % \subfigcapskip = -5pt
	% 				% \subfigure[Yelp]{
		%     %                     \hspace{-5.2mm}
		%     %                     % %\vspace{-1mm}
		% 				% 	\centering
		% 				% 	\includegraphics[width=0.51\linewidth]{img/yelpf1.pdf}
		% 				% } 
	% 				% % \subfigcapskip = -5pt
	% 				% \subfigure[Amazon]{
		%     % 				\hspace{-5.5mm}
		% 				% 	\centering
		% 				% 	\includegraphics[width=0.51\linewidth]{img/bookf1.pdf}
		% 				% }
	%     % \centering
	%     \hspace{-4mm}
	% \includegraphics[width=1.05\linewidth]{img2/flop2.pdf}
	% %\vspace{-0.1in}
	% 				\caption{Performance and computation cost w.r.t. sample rates (portion of the kept items). The dashed line represents the baseline model, i.e., FullSAM.}
	% 				%\vspace{-0.25in}
	% 				\label{flop}
	% 				%\Description{des}
	% 			\end{figure} 
\subsubsection{Efficiency and Effectiveness of Different Relax Factors}
% 		The growing length of the historical behavior sequences results in expensive computation during inference, which greatly increases the burden of online services. 
In practice, it is necessary to balance efficiency and effectiveness when adapting to online services.
Thus, we control the relax factors $b$ in Eq. \ref{rp} to solve this problem. 
% Generally, the smaller $b$ leads to low sample rate which could reduce the computation cost of the sequential recommender.  
The performance and the computation cost of the well-trained model in inference  w.r.t. average sample rate by setting relax factor from $\{-0.5, 0, ... 2 \}$ are shown in Figure \ref{flop}. Note that the computation cost is measured with million floating-point operations (MFLOPs). We surprisingly observe that even preserving 50\% interactions  can achieve the comparable performance as baseline. However, while the sample rate reaches 80\%, the performance becomes to drop which probably because the  sampler leads to a uniform distribution. Besides, the shortened sequences save lots computations especially with small $b$, highlighting the value in practical applications.
% Since the  sampling can be regarded as a binary classification problem which is much simpler than recommendation and requires only one layer,
% %and only needs one layer's computation
% we believe AutoSAM could be more efficient than using full sequence as the model gets deeper. 
\subsubsection{Impact of Sequence Length.}
AutoSAM can model users' non-uniform interest over interacted items, so the method should compatible with different sequence length. To verify such effectiveness, we analyze the performance improvement of AutoSAM on sequences of different lengths.  As shown in Table \ref{seqlen}. We can find that \textbf{(1)} AutoSAM is also effective for users with short sequences. \textbf{(2)} Longer sequences do maximize the effectiveness of sampling, This is probably because longer sequences not only contain richer information but also tend to include noise, offering greater potential for optimization.
\subsubsection{Sampling Quality Evaluation.}	
% Sampling decisions play important roles to the recommend performance while directly affecting the inputs of recommendation models~\citep{ding2020simplify,zhang2021simple}.

\begin{figure}
	\centering
	\hspace{-4mm}
	\vspace{-0.1in}
	\includegraphics[width=0.9\linewidth]{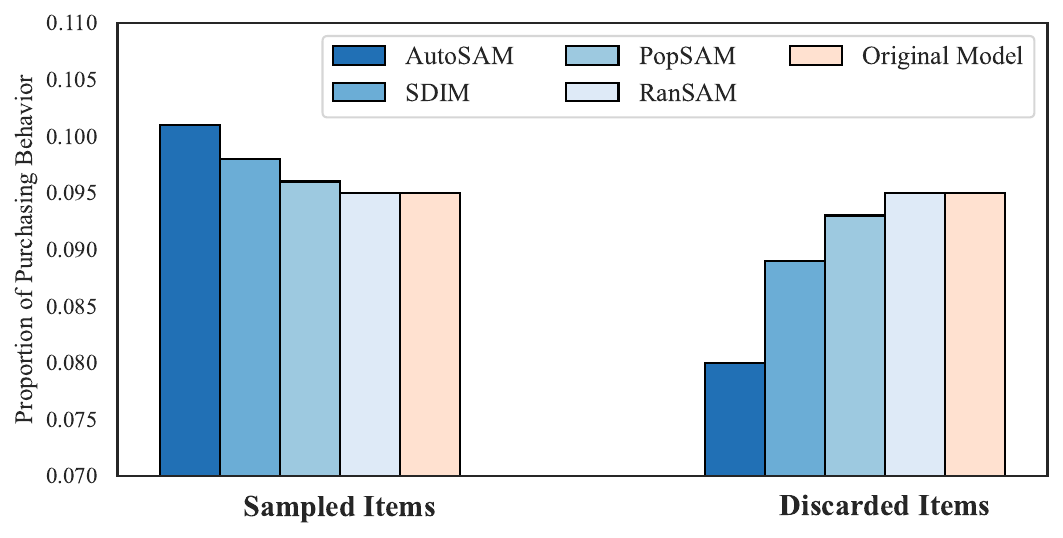}
	\caption{Portion of purchasing behaviors of sampled and discarded interactions on Tmall Dataset.}
	\label{case_p}
	\vspace{-0.2in}
\end{figure}

Next, we evaluate the sample qualities by comparing the portion of purchasing behaviors on sampled and dropped items of different sampling strategy. Commonly, items that are purchased or added to favorites should be of more importance than clicked ones.  As displayed in Figure \ref{case_p}, AutoSAM retains the most purchasing behavior, showing that the learned distributions are more significant to the user's preference, which demonstrates the effectiveness of the sampler and also implies the necessity of treating behaviors non-uniformly.
As it still keeps lots of clicked interactions, a possible reason is that the click items can also reflect the user's interests to a great extent, and the factors behind that are usually very complex in the real world.  
% Finally, we also show several interesting case studies of sampling probability distributions in Appendix.
%  	\begin{figure}[t]
	% % \vspace{-0.09in}
	%        \hspace{-6mm}
	% 				\subfigure[Tmall]{
		%                     \centering
		% 					\includegraphics[width=0.47\columnwidth]{img2/tmall_por3.pdf}
		% 				} 
	% 				\subfigure[Alipay]{
		%                     \centering
		% 					\includegraphics[width=0.47\columnwidth]{img2/ali_por3.pdf}
		% 				}
	%     \vspace{-0.1in}
	% 				\caption{Portion of different kind behaviors of picked and dropped interactions.}
	% 				\label{case_p}
	%     \vspace{-0.25in}
	% 			\end{figure}
%       \begin{figure}[h]
	% 	\centering
	% 	\subfigure[]{
		% 		\centering
		% 	\includegraphics[width=1\linewidth]{img2/case_bar1.pdf}
		% 		%\hspace{5mm}
		% 	}
	% 	\subfigure[]{
		% 		\centering
		%    %\vspace{-0.2in}
		% 		% \hspace{-0.9mm}
		% \includegraphics[width=1\linewidth]{img2/case_bar2.pdf}
		% 	}
	%  %\vspace{-0.1in}
	% 	\caption{Case study analysis in terms of sampling probabilities (P) and book titles on Amazon Dataset.}
	%  %\vspace{-0.2in}
	% 	\label{case}
	%     \end{figure}
\vspace{-0.05in}

\vspace{-0.05in}

\section{Conclusion}
\vspace{-0.05in}
In this work, we proposed a general automatic sampling framework, named AutoSAM, to non-uniformly treat historical behaviors. Specifically, a light-weighted sampler was first leveraged to adaptively explore the distribution of raw input, so that the sequential recommender systems (SRS) could be trained with more informative and diverse samples. Considering the challenges of non-differentiable actions and the necessity to introduce multiple decision factors for sampling,
we further introduced a novel reinforcement learning-based method to guide the training of the sampler
% . We carefully designed multi-objective rewards and combined policy gradient to optimize the whole framework 
in an end-to-end manner. 
We conducted extensive experiments on four public datasets. The experimental results showed that the AutoSAM could obtain a higher performance gain by adaptively sampling informative items.
% In future work, we will explore the incorporation of extra information like ratings or comments to design more comprehensive sampling rewards and extend the proposed framework.
% We also noticed some limitations. AutoSAM conducts a transformer-based sampler, which may not be more efficient than baseline in the shallow SRS, according to Section 3.5. We believe this could be improved by adopting much lighter samplers like RNN or MLP,
% and will study this effect in the future. 
We hope this paper could inspire more works to be proposed from data perspective for SRS.
\\\\
  \textbf{Acknowledgement} This research was supported by grants from the grants of Provincial Natural Science Foundation of Anhui Province (No.2408085QF193), USTC Research Funds of the Double First-Class Initiative (No. YD2150002501), the National Natural Science Foundation of China (62337001), the Key Technologies R \& D Program of Anhui Province (No. 202423k09020039) and the Fundamental Research Funds for the Central Universities (No. WK2150110032).
    
\bibliographystyle{ACM-Reference-Format}
\bibliography{ref}
\end{document}